\documentclass[conference]{IEEEtran}
\IEEEoverridecommandlockouts
\usepackage{amsmath,amssymb,amsfonts}
\usepackage{algorithmic}
\usepackage{graphicx}
\usepackage{textcomp}
\usepackage{xcolor}
\usepackage{booktabs,subcaption,amsfonts,dcolumn}
\usepackage{hyperref}

\def\BibTeX{{\rm B\kern-.05em{\sc i\kern-.025em b}\kern-.08em
    T\kern-.1667em\lower.7ex\hbox{E}\kern-.125emX}}

\begin{document}

\title{Relating Voluntary Turnover with Job Characteristics, Satisfaction and Work Exhaustion -- An Initial Study with Brazilian Developers
\thanks{We appreciate the support of the Federal Police of Brazil (ePol Project) and CAPES.}}

\author{\IEEEauthorblockN{Tiago Massoni}
\IEEEauthorblockN{Nilton Ginani}
\IEEEauthorblockA{\textit{Dep. of Computing and Systems} \\
\textit{UFCG}\\
Campina Grande, Brazil \\
massoni@dsc.ufcg.edu.br\\
nilton.ginani@ccc.ufcg.edu.br}
\and
\IEEEauthorblockN{Wallison Silva}
\IEEEauthorblockN{Zeus Barros}
\IEEEauthorblockA{\textit{Dep. of Computing and Systems} \\
\textit{UFCG}\\
Campina Grande, Brazil \\
wallison.silva@copin.ufcg.edu.br\\
zeus.barros@splab.ufcg.edu.br}
\and
\IEEEauthorblockN{Georgia Moura}
\IEEEauthorblockA{\textit{Dep. of Psychology} \\
\textit{UniNassau}\\
Campina Grande, Brazil\\
georgiaio@hotmail.com}
}
\maketitle
%
\begin{abstract}

High rates of turnover among software developers remain, involving additional costs of hiring and training.
Voluntary turnover may be due to workplace issues or personal career decisions, but it might as well relate to \emph{Job Characteristics}, or even \emph{Job Satisfaction} and \emph{Work Exhaustion}.
This paper reports on an initial study which quantitatively measured those constructs among 78 software developers working in Brazil who left their jobs voluntarily. 
For this, we adapted well-known survey instruments, namely the JDS from Hackman and Oldham's Job Characteristics Model, and Maslach et al.'s Burnout Measurement. 
In average, developers demonstrated low to moderate autonomy ($3.75$, on a 1--7 scale) and satisfaction ($4.08$), in addition to moderate exhaustion ($4.2$) before leaving their jobs, while experiencing high task significance ($5.15$).
Also, testers reported significantly lower job satisfaction than programmers.
These results allow us to raise hypotheses to be addressed by future studies.
\end{abstract}


\section{Introduction}
\label{sec:intro}

As the largest IT market in Latin America --- also World's 9th largest --- Brazil's software industry encompasses around 4,800 companies producing software or providing software-related services, generating, to the local economy, more than US\$8.6 billion (as of 2016~\cite{abes-report}).
In this scenario, high staff turnover becomes critical~\cite{turnover-report}. Software companies face low retention, generating significant costs, due to the time to find other professionals and training new hires~\cite{magazine:lingolive,moore2000one,mcknight2009reduces}.

Similarly to many other areas, software developers may voluntarily leave their current job for many reasons; it is commonly believed the primary motives are better career options or financial improvement, as well as adverse workplace conditions~\cite{mcknight2009reduces}, including communication issues (e.g., programmer-tester conflicts~\cite{test-developer2014}). However, job characteristics (and the developer's emotional response to them) might influence developers in deciding to leave for another company; some of these characteristics may be related to job dissatisfaction and burnout as well~\cite{thatcher-it-worker,robles2006}.
Most studies, such as the one carried out by McKnight et al.~\cite{mcknight2009reduces}, relate job characteristics or exhaustion to \textit{turnover intention}, not actual voluntary turnover. 

In this paper, we describe an initial study with 78 Brazilian software developers who left their last job voluntarily, aiming to explore the relationship between \emph{actual turnover and Job Characteristics}, \emph{Satisfaction} and \emph{Work Exhaustion}.
The survey is primarily based on the \emph{Job Characteristics Theory (JCT)}~\cite{jctOriginal1976} whose primary instrument, the \emph{Job Diagnostic Survey (JDS)}, is adapted to collect data about the respondent's last job, assessing five job core characteristics before they moved to the next (their current) job.
Besides, the survey includes items on Job Satisfaction~\cite{jct-book} and Work Exhaustion (Job Burnout)~\cite{maslach1981}. 
Through mailing lists for software developers, we received 102 answers, from which a sample of 78 left their last job voluntarily. 

Developers reported moderate Work Exhaustion and Job Satisfaction. 
These results agree with research evidence that exhaustion and dissatisfaction are recurrent for technology professionals in general~\cite{moore2000one}.
Also, they often show a lack of autonomy, with 52.6\% of scores lower than 4.
Previous research shows that, in general, autonomy negatively correlates with turnover intention~\cite{mcknight2009reduces,autonomy-correlation}. 
On the other hand, mostly positive scores were observed for Job Significance and Skill Variety.
Significance and Variety seem to be expected by the professional for many development jobs, regardless of which organisation they work for.

As an additional inquiry, we performed comparisons between groups of developers split by demographic information about job position and degree level. No significant difference was observed in most groups; however, our tests detected a slight statistical difference in Job Satisfaction between programmers and testers.

The objective of this initial study is to characterise the job which developers chose to quit voluntarily, aiming at understanding the rationale behind turnover; these results motivate us to carry out, in future work, a comparison of observed measurements between this sample and developers who stayed in the teams.
Also, we raise some hypotheses to be considered in future studies, either with this sample of Brazilian developers or in other contexts. 

\section{Background}
\label{sec:theory}

\noindent\textbf{Developer Turnover.} 
Software companies often face low retention.
Due to the high demand for highly skilled professionals, new jobs are frequently available, and specialised skills generate increasing business costs; companies must find qualified substitutes and train new hires~\cite{mockus2010organizational}.
Turnover is then a significant concern in our software-driven society, having a dramatic impact on project success~\cite{hall2008,hira2016calibrating}. 
Foucault et al.'s work~\cite{foucault2015impact} present evidence that constant human resources change negatively impact software quality. Even reports from open-source projects present a high turnover rate, with dire consequences for project evolution~\cite{hira2016calibrating,Lin2017}.

Academics and industry leaders have tried to address this problem by observing antecedents and consequences of IT workers leaving their jobs voluntarily. 
A systematic review study, for instance, mapped 70 conceptually distinct turnover drivers for those professionals, as categorised into five classes -- individual, job-related, psychological, environmental and organisational factors~\cite{ghapanchi-2011}.
Although incentives like salary and promotion are deemed as critical for leaving jobs, other determinants, such as job autonomy, perceived workload, and satisfaction have their relevance reported by research subjects.  
As an example, Jo Ellen Moore~\cite{moore2000one} assesses a high turnover intention among technology professionals manifesting \emph{work exhaustion}.
Despite its importance, most research relating job-related and psychological factors with turnover focuses on \textit{turnover intention}, in which professionals provide data about the probability of them quitting the current job~\cite{mcknight2009reduces,mockus2010organizational}. 

\noindent\textbf{Job Characteristics Theory.}
In 1975, Oldham and Hackman~\cite{jctOriginal1976} constructed the original version of the Job Characteristics Theory (JCT), to improve work design by measuring the effect of job characteristics on attitudes and behaviours.
In JCT's final version~\cite{jct-book}, five core characteristics (\emph{Skill Variety, Task Identity, Task Significance, Autonomy, and Feedback}) should predict three psychological states (work with meaning, responsibility for results and knowledge of results), leading to favorable personal and work outcomes -- high motivation, satisfaction, and effectiveness.
The numeric relationship between these concepts is conveyed in the \emph{Motivating Potential Score (MPS)}, an index of the ``degree to which a job (...) is likely to prompt favourable personal and work outcomes''~\cite{jctOriginal1976}.
The Job Diagnostic Survey (JDS) is employed to assess JCT, measuring jobholders' perceptions through 17 Likert-based items.

Nearly JDS 200 studies have been meta-analytically reviewed~\cite{fried1987validity,deVaroJCT}, indicating the correlational results are reasonably valid. 
The model has endured criticism, even though it is still recognised as useful for assessing motivation and satisfaction at work, including IT-related jobs~\cite{deMagalhaes-2017,mcknight2009reduces}.

\noindent\textbf{Work Exhaustion.}
Maslach and Jackson~\cite{maslach1981} define \emph{work exhaustion} (job burnout) as a psychological syndrome of emotional exhaustion, depersonalization (negative or detached behaviour toward others), and diminished personal accomplishment. 
Many careers may experience burnout, but IT professionals can be particularly susceptible; several studies suggest the prevalence of work overload~\cite{Sethi-1999,cook2015job}. 

\section{Method}
\label{sec:method}

We structure the study in terms of two research questions:

\textit{RQ1: What are the perceived job characteristics, burndown, and satisfaction in the job they voluntarily quit?} We aim to assess the average scores for each JCT core characteristic (Skill Variety, Task Identity, Job Significance, Autonomy, Job Feedback), Work Exhaustion and Job Satisfaction.

\textit{RQ2: How do those values relate to Job Position and Degree Level?} We analyse the scores as compared with demographic data: Job Position (Programmer, Tester, and Manager), and Degree Level (Undergraduate, Graduate, and M.Sc. or Ph.D.).

\subsection{Study Participants}

Participants are software developers, working either in public or private companies in Brazil, who must have worked in at least two paid jobs; all items apply only to their previous job.
Also, they were asked to specify their position within the team. 

We sent the invitation with an access link for an online form with informed consent~\footnote{All study's material and data are available as a \href{https://doi.org/10.5281/zenodo.2552240}{Zenodo repository}~\cite{rep-package}.} to two online groups of software developers, four mailing lists, and to about 40 directed people, who served as a hub to pass on the survey to colleagues, as a convenience sample.
The survey remained open to collect answers from December 15th, 2017 to February 12th, 2018.

\subsection{Procedure and Measurement}

The JDS instrument covers five job core characteristics, encompassing 17 items, while Job Satisfaction is measured with three Likert-scale items~\cite{jct-book}. 
Work Exhaustion is assessed with four items, as adapted from Maslach et al.'s Burnout Measurement~\cite{maslach1981}, asking about feelings of exhaustion before and after a day at the job, or during job tasks
~\cite{moore2000one,mcknight2009reduces}.
Question ordering was random, to minimise grouping bias. 

The survey was applied, at first, to a small sample of 15 participants, who were asked to provide feedback about the items. 
As a measure of internal consistency (how related each set of items is into a group), we calculated Cronbach's $\alpha$, which showed values of at least $0.87$ -- with $\alpha > 0.8$, an instrument is considered better than acceptable. 
After fixing issues of understandability (question phrasing from the translation to Portuguese), we carried out the final study.
For the JCT core characteristics, we calculate, for each participant, the Motivating Potential Score (MPS), which is given as follows: $MPS = \frac{SkillVar+TaskId+TaskSig}{3} \times Auton \times Feedb$. 
We used RStudio~\cite{rstudio} as the analytical tool for the R language. 

The study has inherent limitations.
First, as an external threat, from a sample of 78 Brazilian developers results cannot be generalizable, even though we reach a diverse range of software companies in different regions in Brazil by using the online survey.
Also, scores might have been affected by the time passed from when participants left by the previous company, or even for justifying their decision if compared to their current position.
We intend to isolate this time factor in our future qualitative studies with the sample.

\section{Results and Discussion}
\label{sec:results}

\subsection{Research Question 1}


Table~\ref{tab:resultsJCT} presents average scales and standard deviation for each JCT core characteristic, along with the average MPS.
If MPS scores are classified as low (for scores below $50$), moderate ($50$--$87.5$) and high ($>87.5$), the average MPS ($81.31$) is within a 95\%-significance confidence interval of [$67.16$,$95.46$] --  moderate to high.

\begin{center}
\begin{table}
\caption{Average scores.}
\begin{center}
\begin{subtable}[t]{0.4\textwidth}
\begin{tabular}[t]{@{}lcrr@{}}
\toprule
\textbf{JCT Core Char.} & \multicolumn{1}{c}{\textbf{Items}} & \multicolumn{1}{r}{\textbf{Average}} & \multicolumn{1}{r}{\textbf{Stand. Dev.}} \\ \midrule
Skill Variety & 4 & 4.68 & 1.70 \\
Task Identity & 3 & 4.29 & 1.49 \\
Task Significance & 3 & 5.15 & 1.41 \\
Autonomy & 4 & 3.75 & 1.49 \\
Feedback & 3 & 4.16 & 1.51 \\ \hline
\textbf{MPS} & \textbf{-} & \textbf{81.31} & \textbf{62.75} \\ \bottomrule
\end{tabular}
\caption{JCT Results}
\label{tab:resultsJCT}
\end{subtable}
\end{center}
\begin{center}
\begin{subtable}[t]{0.4\textwidth}
\begin{tabular}[t]{@{}lcrr@{}}
\toprule
\textbf{Concept} & \multicolumn{1}{c}{\textbf{Items}} & \multicolumn{1}{r}{\textbf{Average}} & \multicolumn{1}{r}{\textbf{Stand. Dev.}} \\ \midrule
Work Exhaustion & 4 & 4.0 & 1.63 \\
Job Satisfaction & 3 & 4.08 & 1.48 \\ \bottomrule
\end{tabular}
\caption{Satisfaction and Burndown Results}
\label{tab:satburn}
\end{subtable}
\end{center}
\end{table}
\end{center}

Autonomy received the lowest scores regarding the developers' last job, from which they left voluntarily.
Considered as the desire to be self-directed, which is related to independence in work, inferior assessments were reported by most participants (52.6\% of answers below $4$). 
This result suggests most professionals lacked autonomy in the job they were willing to leave. Previous research work shows that, in general, autonomy negatively correlates with turnover intention~\cite{mcknight2009reduces,autonomy-correlation}.
Furthermore, most professionals seemed to receive moderate feedback from their tasks in their last job -- software professionals are reported to benefit from honest and throughout feedback~\cite{feedback-importance}. 
On the other hand, at least two JCT core characteristics were positively assessed, on average: Job Significance and Skill Variety. Previous research has found a weak relationship between those aspects and turnover intention~\cite{mcknight2009reduces}.


Regarding Job Satisfaction, Table~\ref{tab:satburn} shows a moderate average score from its three items, $4.08$, with low standard deviation ($1.48$), suggesting developers, although not reporting a significantly negative experience, either did not feel inspired or stimulated by the job.
Lower scores usually convey disapproval on the way they worked, with the position they held, or the company they worked for. 
Data indicate less satisfied developers tend to look for another job. This is reinforced by the high demand for software developers, making them less cautious about their current job when looking for a more satisfying position.

In terms of Work Exhaustion, 25\% of the participants answered they felt emotionally exhausted at least once a week (9\% reported they felt like this \textit{every day}), 29\% reported feeling physically exhausted in a weekly basis, and half of the participants reported feeling tired in the morning, before starting a work day in their last job.
Work exhaustion is linked to psychiatric disorders such as the burnout syndrome, hampering motivation for work, often making professionals to look for a potentially less stressful job. There is evidence that work exhaustion is recurrent for technology professionals in general~\cite{moore2000one}.

\subsection{Research Question 2}

Regarding Job Position, open-ended space-limited answers were manually classified as a programming position (66.7\%), testing position (21.8\%) and management position (11.5\%). 
Regarding Degree Level, around 7\% are professionals with no diploma yet, and more than 59\% are at least graduated. Almost one-third of the sample holds M.Sc. or Ph.D. degrees.

We found no significant difference in any measurement when it comes to Degree Level (Table~\ref{tab:kruskalLevel}).
Besides, we compared average MPS, Job Satisfaction and Work Exhaustion scores between the three groups of job positions.
For MPS and Exhaustion, difference results do not present statistical significance (the null hypotheses with corresponding p-values are also shown in Table~\ref{tab:kruskalPos}, by using Kruskal-Wallis variance tests~\cite{Kruskal}). 
On the other hand, the test rejects the null hypothesis for Job Satisfaction (assuming $95\%$ as confidence level), suggesting a difference among the groups for \emph{satisfaction}; by applying a Tukey \emph{post-hoc} test, we see a slightly significant difference in satisfaction between programmers and testers (with adjusted $p-value=0.049$).
Research evidence~\cite{zhang2010organizing} suggests quality assurance activities could be less challenging than design and programming tasks, which could be a potential explanation for such results.

\begin{table}
\caption{Hypothesis test results.}
\begin{center}
\begin{subtable}[t]{0.4\textwidth}
\centering
\begin{tabular}[t]{lr}
\toprule
 \textbf{Null Hypothesis} & \textbf{Kruskal-Wallis}\\
 & \multicolumn{1}{c}{\textbf{p-value}} \\ 
 \hline
 MPS & 0.59 \\ 
 \textbf{Job Satisfaction} & {\bfseries0.03} \\ 
 Work Exhaustion & 0.25 \\ 
 \bottomrule
\end{tabular}
\caption{Job Positions.} 
\label{tab:kruskalPos}
\end{subtable}
\end{center}

\begin{center}
\begin{subtable}[t]{0.4\textwidth}
\centering
\begin{tabular}[t]{lr}
\toprule
 \textbf{Null Hypothesis} & \textbf{Kruskal-Wallis}\\
 & \multicolumn{1}{c}{\textbf{p-value}} \\ 
 \hline
 MPS & 0.16 \\ 
 Job Satisfaction & 0.16 \\ 
 Work Exhaustion & 0.14 \\ 
 \bottomrule
\end{tabular}
\caption{Degree Levels.} 
\label{tab:kruskalLevel}
\end{subtable}
\end{center}

\end{table}

\section{Conclusion}
\label{sec:conclusion}

In this paper, we performed an initial study on assessing Job Characteristics, Satisfaction and Work Exhaustion from software developers who left their last job voluntarily. Seventy-eight developers located in Brazil responded to the survey, whose instrument was adapted from the JSD and Maslach et al.'s Burnout Measurement~\cite{maslach1981}. As the next step, we intend to compare these results with more recently developed survey instruments related to job characteristics, like the Work Design Questionnaire (WDQ)
~\cite{morgeson2003work}.
Some hypotheses could be further investigated: despite a positive perception about the variance of skills or significance of the produced software, developers might look for other jobs due to other factors.
We intend to carry out qualitative studies with the same participants, to further inquire into the reasons for the job change.
Similarly, Job Satisfaction was perceived as moderate to low for almost 40\%  of the sample, suggesting a bad feeling or negative emotions were noticeable from their last job experience.

Furthermore, we investigated the relationship between the measured constructs and demographic data, namely Job Position and Degree Level. No significant differences were observed in most groups; 
only a slight statistical difference between programmers and testers, regarding Job Satisfaction, was detected. 
This hypothesis may be subject to further research: are Quality Assurance (QA) professionals less satisfied with their jobs, if compared to programmers? Or, if this is true, are they leaving their QA jobs for programming positions?

While only a few studies have examined motivation with actual turnover, it is hoped that additional studies of software developers will be conducted to more firmly determine the boundaries of generalisability. 
As further work, for instance, we intend to investigate developers who still work with the teams left by our participants, giving the means for comparing the measured results with a baseline. 
Scientific evidence on this matter could guide companies in avoiding the loss of key developers and its damaging consequences.

\bibliographystyle{IEEEtran}

\end{document}